\documentclass[conference,a4paper,twocolumns]{IEEEtran}
\IEEEoverridecommandlockouts

\usepackage{amsmath, amssymb, cite}
\usepackage{amsthm,bm,bbm}
\usepackage{mathtools}
\usepackage[size=small]{caption}
\usepackage{amsthm,multirow,color,amsfonts}
\usepackage{tabulary}
\usepackage{subfigure}

\usepackage{graphicx}
\usepackage{setspace}
\usepackage{enumerate}
\usepackage{algorithm}
\usepackage{algpseudocode}
\usepackage{comment}
\usepackage{pdfpages}

\usepackage{etoolbox}
\makeatletter
\patchcmd{\@makecaption}
  {\scshape}
  {}
  {}
  {}
\makeatother

\makeatletter
\def\BState{\State\hskip-\ALG@thistlm}
\makeatother

\usepackage[utf8]{inputenc} 
\usepackage[T1]{fontenc}
\usepackage{ifthen}

\usepackage[dvipsnames]{xcolor}

\newtheorem{theo}{Theorem}
\newtheorem{defi}{Definition}

\newtheorem{cor}{Corollary}

\newtheorem{lem}{Lemma}
\newtheorem{ex}{Example}

\usepackage{mathtools}
\usepackage{multirow,color,amsfonts}
\usepackage{tabulary}
\usepackage{subfigure}
\usepackage{graphicx}
\DeclareGraphicsExtensions{.png,.pdf}
\usepackage{setspace}
\usepackage{enumerate}
\usepackage{comment}

\usepackage[utf8]{inputenc} 
\usepackage[T1]{fontenc}
\usepackage{ifthen}

\usepackage[hyphens]{url}
\usepackage[hidelinks]{hyperref}
\hypersetup{breaklinks=true}
\usepackage{tikz}

\usepackage{etoolbox}
\makeatletter\patchcmd{\@makecaption}  {\scshape}  {}  {}  {}\makeatother

\usepackage{amsthm}
\usepackage{amsmath}

\def\compactify{\itemsep=0pt \topsep=0pt \partopsep=0pt \parsep=0pt}
\let\latexusecounter=\usecounter

\begin{document}
\title{The Influence of Placement on Transmission in Distributed Computing of Boolean Functions}


 \author{
   \IEEEauthorblockN{Ahmad Tanha,~\IEEEmembership{Student Member,~IEEE}, Derya Malak,~\IEEEmembership{Member,~IEEE}}
   \IEEEauthorblockA{Communication Systems Department, EURECOM, Sophia Antipolis, France\\
   \{ahmad.tanha, derya.malak\}@eurecom.fr}
\thanks{Funded 
by the European Union (ERC, SENSIBILITÉ, 101077361). Views and opinions expressed are however those of the authors only and do not necessarily reflect those of the European Union or the European Research Council. Neither the European Union nor the granting authority can be held responsible for them. 

This research was partially supported by a Huawei France-funded Chair towards Future Wireless Networks, and supported by the program ``PEPR Networks of the Future" of France 2030. }\vspace*{-0.75cm}
}

\maketitle

\thispagestyle{plain}
\pagestyle{plain}

\begin{abstract}
In this paper, we explore a distributed setting, where a user seeks to compute a linearly-separable Boolean function of degree $M$ from $N$ servers, each with a cache size $M$. Exploiting the fundamental concepts of sensitivity and influences of Boolean functions, we devise a novel approach to capture the interplay between dataset placement across servers and server transmissions and to determine the optimal solution for dataset placement that minimizes the communication cost. In particular, we showcase the achievability of the minimum average joint sensitivity, $\frac{N}{2^{M-1}}$, as a measure for the communication cost.
\end{abstract}


\section{Introduction}
\label{sec:intro}
Over the past few decades, technological advancements have significantly increased the demand for high-performance distributed computing to divide a computationally heavy task into multiple subtasks with lower computation load over workers across a network, e.g., 
machine learning algorithms over distributed servers \cite{Li2023ISIT}, and 
cloud computing platforms 
\cite{Xu2018}. Even though there exist heuristic approaches to the problem of distributed computing in the literature, such as MapReduce \cite{mapred2008}, 
managing the ever-increasing demands requires a deep understanding of distributed placement, compression, and transmission of datasets towards realizing various computations, which is our key focus in this paper.

\subsection{Related Work}
We first review 
functional compression literature. We then provide the existing algorithms for distributed placement of datasets to achieve various computation tasks.


\paragraph{From source compression to functional compression}
While the fundamental limits for the problem of data compression, either centralized \cite{shannon1948mathematical} or distributed \cite{slepian1973noiseless}, has been explored, the general problem of compression for computing, or functional compression, requires different tools that can exploit the structure of the computation task. 
To that end, K\"orner introduced the notion of graph entropy for distinguishing source symbols that produce different function outcomes \cite{korner1973coding}, and the concept of graph coloring was later used in various distributed functional compression settings, including but not limited to \cite{OrlRoc2001,FM14,Malak2023SPAWC}. 
However, this technique may not apply to general functions. 
Other works, e.g., \cite{korner1979encode,han1987dichotomy}, to exploit the characteristics of functions, devised structured coding schemes, which require different encoding functions for different tasks and hence may not be practical.
 


Following the coded distributed computing scheme in \cite{Li2018TIT}, several works investigated the storage-computation-communication tradeoffs, e.g., for the class of linearly-separable functions, using cyclic placement \cite{Wan2022TIT2}, or with linear coding for optimizing placement and transmissions \cite{Khalesi2023TIT}, using placement delivery arrays \cite{Yan2022TIT}, and tessellations \cite{khalesi2024tessellated}.
\paragraph{The role of placement in distributed computation} 
The most common placement of datasets across servers in distributed setting's literature is the ``cyclic placement'' scheme on datasets, e.g., as in \cite{Wan2022TIT2}.
The placement on distributed servers is conducted in a cyclic manner, in the amount of some circular shifts between two consecutive servers.
As a result of cyclic placement, any subset of servers covers the set of all datasets to compute the requested functions from the user.

While the main focus in the related literature is on encoding, transmission, and decoding phases with a given assignment of data, the placement configuration of datasets across servers could significantly affect the communication cost \cite{Khalesi2023TIT, khalesi2024tessellated}. 




\subsection{Motivation and Contributions}
Motivated by the impact of dataset placement on transmission in distributed computing systems \cite{Khalesi2023TIT, khalesi2024tessellated},
we utilize the concept of sensitivity and influence of Boolean functions \cite{bourgain1992influence}, \cite{jukna2012boolean} to designate an optimal placement configuration that achieves the minimum communication cost.
Specifically, we focus on the linearly-separable Boolean functions. 

In this paper, we present a novel distributed computing approach that involves a master node, a set of distributed servers, and a user demanding the error-free computation of a linearly-separable Boolean function. The master node distributes datasets across servers, where each server then performs subcomputations of datasets. 
Our approach captures the joint influences of subsets of distributed datasets in computing the user demand for any given number of servers with identical cache sizes. 
This enables us to show the fundamental interplay between the placement and transmission for distributed computation of linearly-separable Boolean functions, where the function structure reveals an optimal placement configuration.


\subsection{Organization}
The rest of the paper is organized as follows. In Section~\ref{sec:model}, we present the proposed scheme for distributed computing of a Boolean function, and the relation between dataset placement and transmissions. 
Next, in Section~\ref{sec:main_results},  exploiting the notions of sensitivity and influences of Boolean functions given placement configurations, we propose a novel approach for analyzing the communication cost for distributed computing of linearly separable Boolean functions.  
Finally, in Section~\ref{sec:conclusion}, we discuss potential future directions toward extending the influence-based concept to a general class of functions.


{\bf Notation.} 
The notation $\mathbb{F}_2^L$ represents the binary field of length $L$, where $\mathbb{F}_2=\{0,1\}$. 
We use square brackets to represent a set of integers, where $[K]\triangleq\{1,2,\dots,K\}$, given $K\in\mathbb{Z}^+$, and curly brackets to denote a set of subsets, e.g., $\{\mathcal{S}_n\}$, where $\mathcal{S}_n$ is a subset of datasets. For a random variable $X$, $\mathbb{E}[X]$ is its expected value.
We denote by $\mathbf{W}=(W_1,\ldots, W_K)$ the vector of all datasets. The basis vector notation 
$e_k\triangleq(0,\dots, 0,1,0,\dots,0)$ represents a binary vector with cardinality $K$ such that $e_k(k)=1$ and $e_k(l)=0$, $\forall$ $l\neq k$.
The notations $\oplus$ and $\bigoplus$ indicate the modulo two addition and the summation symbol 
in $\mathbb{F}_2$, respectively. 
Hence, $\mathbf{w}\oplus e_k$ represents $\mathbf{w}$ with the $k^{th}$ entry flipped. We denote the indicator function by $1_{\{.\}}$. 



\section{System Model}
\label{sec:model}

We consider a distributed computing setup consisting of a master node, a set of distributed servers, and a user. 
In this setting, there are $K$ independent and identically distributed (i.i.d.) datasets, where each dataset $k\in [K]$ is a Bernoulli distributed random variable, denoted by $W_k\sim {\rm Bern}(\frac{1}{2})$. The master node assigns (possibly not disjoint) subsets of datasets 
$N$ servers indexed by $[N]$, where each server has an identical cache size that allows storing up to $M$ datasets. Finally, the user seeks to compute a Boolean function\footnote{A function $f$ is called Boolean if it only accepts binary values as both domain and range, i.e., $0$ and $1$ \cite{jukna2012boolean}.} $f:\mathbb{F}_2^K\longrightarrow \mathbb{F}_2$ of the input vector of all $K$ datasets, i.e., $\mathbf{W}=(W_1, W_2,\ldots, W_k)$.

In this paper, we use the below representation for a Boolean function in general polynomial form \cite{jukna2012boolean}:
\begin{align}
\label{def:Boolean_function}
 f(\mathbf{W})= \bigoplus_{\mathcal{P}\subseteq [K]} c_{\mathcal{P}} \prod_{k\in \mathcal{P}} W_k
\end{align}
for some subsets $\mathcal{P}$ of $K$ datasets and coefficients $c_{\mathcal{P}}\in \mathbb{F}_2$. 

\subsection{Phases of Distributed Computing}
\label{sec:placement_transmission_decoding}
In this setting, we have three phases for distributed computing of $f(\mathbf{W})$ given the input vector $\mathbf{W}$, as described next.

\paragraph{Dataset Placement}
In this phase, the master node will assign a subset of datasets to each server node without coding across different datasets, known as uncoded placement in the literature, see e.g., \cite{Jin18}. In particular, given a cache size of $M$ for each server, the master node will assign subsets of $K$ datasets to the servers according to a placement function $\rho_n$, $n\in [N]$ which is described as
\begin{align}
    \rho_n: \mathbb{F}_2^K\longrightarrow \mathbb{F}_2^M \ , \quad\forall n\in [N] \ ,
\end{align}
where the assigned subset to the $n^{th}$ server is specified as 
\begin{align}
    S_n=\rho_n(\mathbf{W})\subseteq \{\mathbf{W}\}\ ,\quad |S_n|=M \ ,\quad\forall n\in [N] \ .
\end{align} 
In other words, the placement phase assigns subsets with a cardinality $M$ of possibly overlapping datasets to the servers. 
We also denote the set of the assigned subsets, i.e., placement configuration by $\mathcal{S}=\{S_n\,|\,n\in [N]\}$.

\paragraph{Encoding and Transmissions}
Given the subsets $S_n, n\in[N]$ from the placement phase, we next detail the encoding phase. 
Given the subset $S_n$ and the computation task $f(\mathbf{W})$, server $n\in [N]$ will conduct subcomputations to determine its transmitted information. Here we note that the servers know the task $f(\mathbf{W})$ a priori and design the subcomputations accordingly.  
We model the encoding and transmission process at server $n\in [N]$ by the function 
\begin{align}
    E_n^f: \mathbb{F}_2^M\longrightarrow \mathbb{F}_2^{|Z_{n}|} \ ,
\end{align}
where we denote the set of computed data by server $n$ by 
\begin{align}
Z_n=E_n^f(S_n)=\{Z_{ni}\,|\,i\in [|Z_n|]\}\ , \quad n\in[N] 
\end{align}
which is then transmitted to the user. 
We also denote by $Z=\{Z_n\,|\,n\in [N]\}$ the set of all transmitted data by all servers to measure the total number of transmissions. 
The user, as we describe next, will aggregate the transmissions $Z$ from all servers to determine the output of $f(\mathbf{W})$.

\paragraph{Decoding}
We assume that for any given placement configuration and given input vector $\mathbf{W}$, once the user receives the subcomputations from each server, it will be able to calculate the outcome of the Boolean function, i.e., $f(\mathbf{W})$, represented by the general polynomial form in (\ref{def:Boolean_function}). The decoding procedure should be designed based on the placement scheme and the encoding process. The decoding function for the recovery of the function by the user is specified as 
\begin{align}
    D:\mathbb{F}_2^{|Z|}\longrightarrow \mathbb{F}_2 \ .
\end{align} 

We assume that the user can recover the function without any error. Hence, for error-free recovery of the 
computation task,  the decoding procedure must satisfy $D(Z)=f$.

\begin{figure}[t!]
    \begin{center}
        \includegraphics[width=0.4\textwidth]{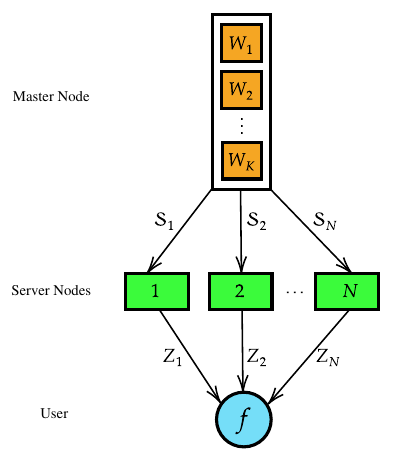}
    \end{center}
    \vspace{-0.3cm}
    \caption{\small A generic distributed computing system model.}
    \label{distcomp}
\end{figure}

We illustrate the system model for our distributed computing scheme in Figure~\ref{distcomp}. 
In this work, we focus on linearly-separable Boolean function of degree $d$ as 
\begin{align}
\label{lin_sep}
   &f(\mathbf{W})=\bigoplus_{n=1}^N f_n \ ,\nonumber\\
   &f_n=\underset{k\in\mathcal{K}_{\mathcal{P}_n, d}}{\prod} W_k\ ,
\end{align}
where $\mathcal{K}_{\mathcal{P},d}$ is a subset $\mathcal{P}$ with cardinality $d$ of $K$ datasets:
\begin{align}
\label{eq:KPd}
\mathcal{K}_{\mathcal{P},d}\triangleq\{ \mathcal{P}\subseteq{[K]} \ \vert \ |\mathcal{P}|=d\}\ ,
\end{align}
where $d=M$. We also assume that $K=NM$, implying $N$ servers each with cache size $M$. 

We refer to our system model as a $(K, N, M, \mathcal{S}, f)$ distributed computing scheme. 
We next define an achievable scheme for error-free distributed computing of $f$ at the user.

\begin{defi}({\bf An achievable distributed computing scheme.})
\label{ach}
A $(K, N, M, \mathcal{S}, f)$ distributed computing scheme is called achievable if the function $f$ can be recovered in an error-free manner by the user with the given cache configuration, i.e., $D(Z)=f$, where $Z$ is possibly a nonlinear combination of the encoded data $Z_n=E_n^f(\mathcal{S}_n)$, $n\in [N]$, which is placement-dependent and function-aware.
\end{defi}

Exploiting   
the definition of $Z=\{Z_n\,|\,n\in [N]\}$, we denote the total number of transmissions by all servers as 
\begin{align}
\label{eq:no_of_transmissions}
T^{(\mathcal{S})}\triangleq |Z|=\sum_{n=1}^N |Z_n| \ .
\end{align}

We next consider an example to demonstrate the interplay between the placement configuration and the value of $T^{(\mathcal{S})}$ for the given $(K, N, M, \mathcal{S}, f)$ distributed computing scheme. 
We will then show the connection between our model for communication cost and $T^{(\mathcal{S})}$ 
in Section~\ref{sec:main_results}.


\subsection{The Interplay between Placement and Transmission}
\label{sec:example}
In this subsection, we first present an example, with two different placement configurations, namely $\mathcal{S}^{(1)}$ which is cyclic, and $\mathcal{S}^{(2)}$, for computing a Boolean function. We next contrast the total number of transmissions needed in each configuration to demonstrate the role of placement and transmissions.

\begin{ex}
\label{ex1}
Consider a $(K=9,\ N=3,\ M=6,\ \mathcal{S},\ f)$ distributed computing system, where 
\begin{align}
\label{Boolean_function_example}
f(\mathbf{W})=W_1 W_4 W_7\oplus W_2 W_5 W_8 W_7\oplus W_3 W_6 W_9 \ .
\end{align}

\begin{enumerate}[(i)]
    \item {\bf Configuration $\mathcal{S}^{(1)}:$}     
    The dataset assignment is cyclic such that $\mathcal{S}^{(1)}_1=\{W_1,W_2,W_3,W_4,W_5,W_6\}$, $\mathcal{S}^{(1)}_2=\{W_4,W_5,W_6,W_7,W_8,W_9\}$, and $\mathcal{S}^{(1)}_3=\{W_1,W_2,W_3,W_7,W_8,W_9\}$, satisfying the cache size constraint with equality.  
    To successfully compute (\ref{Boolean_function_example}) in this configuration, it suffices that the servers need to compute and transmit data as shown in Table~\ref{tab:t1}. 
    Hence, this scenario requires $T^{\mathcal{S}^{(1)}}=6$ transmissions in total.
    
    \begin{table}[t!]\small
        \begin{center}
        \tabcolsep=0.11cm
        \begin{tabular}{c|c}
           Assigned subsets & Transmitted data\\
           \hline
            $S_1^{(1)}$ & $Z_{11}^{(1)}=W_1W_4$, $Z_{12}^{(1)}=W_3W_6$, $Z_{13}^{(1)}=W_2$\\
            \hline
            $S_2^{(1)}$ & $Z_{21}^{(1)}=W_5W_8W_7$, $Z_{22}^{(1)}=W_7$, $Z_{23}^{(1)}=W_9$\\
            \hline
            $S_3^{(1)}$ & No transmissions 
        \end{tabular}
        \end{center}
       \caption{\small Server-transmission details for $\mathcal{S}^{(1)}$.}
        \label{tab:t1}
    \end{table}
   
    \item {\bf Configuration $\mathcal{S}^{(2)}:$} 
    The dataset assignment satisfies $\mathcal{S}^{(2)}_1=\{W_1,W_2,W_4,W_5,W_6,W_7\}$, $\mathcal{S}^{(2)}_2=\{W_3,W_4,W_5,W_6,W_8,W_9\}$, and $\mathcal{S}^{(2)}_3=\{W_1,W_2,W_3,W_7,W_8,W_9\}$. 
    To successfully compute (\ref{Boolean_function_example}) in this setting, without the presence of stragglers, a viable transmission scheme is shown in Table~\ref{tab:t2}. 
    Thanks to a better arrangement of the datasets that is sensitive to the function in $\mathcal{S}^{(2)}$ versus $\mathcal{S}^{(1)}$, the total number of transmissions is $T^{\mathcal{S}^{(2)}}=4<T^{\mathcal{S}^{(1)}}=6$.
    
    \begin{table}[t!]\small
        \begin{center}
        \tabcolsep=0.11cm
        \begin{tabular}{c|c}
           Assigned subsets & Transmitted data\\
           \hline
            $S_1^{(2)}$ & $Z_{11}^{(2)}=W_1W_4W_7$, $Z_{12}^{(2)}=W_2W_5W_7$\\
            \hline
            $S_2^{(2)}$ & $Z_{21}^{(2)}=W_3W_6W_9$, $Z_{22}^{(2)}=W_8$\\
            \hline
            $S_3^{(2)}$ & No transmissions 
        \end{tabular}
        \end{center}
        \caption{\small Server-transmission details for $\mathcal{S}^{(2)}$.}
        \label{tab:t2}
    \end{table}
    
     
\end{enumerate}
\end{ex}


We infer from Example \ref{ex1} how a cleverly conducted placement phase in distributed computing settings could dramatically reduce the total number of transmissions needed for error-free recovery of the Boolean function at the user.

To evaluate the communication cost of the $(K, N, M, \mathcal{S}, f)$ distributed computing scheme, we will next detail a novel approach that relies on the average joint sensitivity of the computation task abstracted by the Boolean function.

\section{Main Results}
\label{sec:main_results}
In this section, to determine the role of the placement configuration on the communication cost, we first provide a primer on the sensitivity of a Boolean function on its input, and the influence of a set of input variables on the outcome of the function. We then present our main results.

\subsection{Joint Sensitivity and Influences}
\label{sec:influ}

Building on the classical notions of sensitivity, influence, and average sensitivity tailored for capturing the sensitivity of a Boolean function by modifying one dataset at each time \cite{bourgain1992influence,jukna2012boolean}, we will exploit the joint behavior of datasets across servers, as in \cite{Sarkar2023}. 
A Boolean function $f(\mathbf{W})$ 
depends on its $k^{th}$ input variable if there exists at least one $\mathbf{W}\in \mathbb{F}_2^K$ such that $f(\mathbf{W}\oplus e_k)\neq f(\mathbf{W})$.
To that end, we next define the sensitivity of $f(\mathbf{W})$ on a set of $W_k$'s.

\begin{defi}{\bf (Joint sensitivity.)} 
\label{joint_sen}
The joint sensitivity of $f(\mathbf{W})$ to the set $\mathcal{S}$ of subsets of datasets is  defined as
\begin{align}
    {\rm Sen}_{\mathcal{S}}(f,\mathbf{W})=\sum\limits_{n=1}^{|\mathcal{S}|} {\rm Sen}_{\mathcal{S}_n}(f,\mathbf{W}) \ , 
\end{align}
where 
$e_{\mathcal{S}_n}\triangleq\underset{\{k | w_k\in \mathcal{S}_n\}}{\bigoplus} e_k$ describes the multi-dataset flipped vector,
the measure ${\rm Sen}_{\mathcal{S}_n}(f,\mathbf{W})=1_{f(\mathbf{W}\oplus e_{\mathcal{S}_n})\neq f(\mathbf{W})}$ captures the sensitivity of $f(\mathbf{W})$ on input $\mathbf{W}$ with the jointly flipped entry datasets with indices $k$ such that $W_k\in \mathcal{S}_n\subseteq \mathcal{S}$. 

\end{defi}

\begin{defi}{\bf (Joint influence.)}
\label{joint_inf}
The joint influence of the datasets of 
$\mathcal{S}_n$ on the function $f$ is defined as
\begin{align}
    {\rm Inf}_{\mathcal{S}_n}(f)&=\mathbb{P}[f(\mathbf{W}\oplus e_{\mathcal{S}_n})\neq f(\mathbf{W})]\nonumber\\
    &=\mathbb{E}_{\mathbf{W}}[{\rm Sen}_{\mathcal{S}_n}(f,\mathbf{W})]\ .
\end{align}
\end{defi}

Definitions~\ref{joint_sen}-\ref{joint_inf} allow us to introduce our next key metric.

\begin{defi}{\bf (Average joint sensitivity.)}
\label{def:ajs}
The average joint sensitivity of $f(\mathbf{W})$ 
to the set $\mathcal{S}=\{S_n | n\in[N]\}$ of all possible datasets specified by $\rho_n$ given a cache size constraint $M$ is given as follows:
\begin{align}
{\rm as_{\mathcal{S}}}(f)&=\mathbb{E}_{\mathbf{W}}[{\rm Sen}_{\mathcal{S}}(f,\mathbf{W})]\nonumber\\
&=\sum\limits_{n=1}^{|\mathcal{S}|} {\rm Inf}_{\mathcal{S}_n}(f)\ . 
\end{align}
\end{defi}

Utilizing Definitions \ref{joint_sen}-\ref{def:ajs}, we can next present our approach for formulating an optimal dataset placement configuration from a communication cost perspective.

\subsection{The Communication-Optimal Placement Configuration}
\label{sec:opt}

To evaluate the tradeoff between placement and communication cost for computing a Boolean function $f$, we first present a Lemma that focuses on each product subfunction $f_n$ 
given in \eqref{lin_sep} to obtain the joint influence of datasets on $f_n$.

\begin{lem} {\bf (Joint influence on a product subfunction.)}
\label{lem_inf_prod}
The joint influence of multiple datasets in a subset $\mathcal{S}_n$ with an arbitrary size from a product subfunction $f_n(\mathbf{W})=\underset{{k\in \mathcal{K}_{\mathcal{P},d}}}{\prod} W_k$ of degree $d$ equals the influence of each dataset on $f_n$, i.e., 
\begin{align}
\label{eq_inf_prod}
    {\rm Inf}_{\mathcal{S}_n}(f)={\rm Inf}_{k}(f)=\frac{1}{2^{d-1}}\ .
\end{align}

 \begin{proof}
     See Appendix \ref{App_A}.
 \end{proof}   
    
\end{lem}

Towards determining the joint influence of datasets on $f$ in \eqref{lin_sep}, we next present another Lemma that contrasts the joint influence of datasets for the summation of two product subfunctions, namely $f_n$ and $f_{n'}$ where $n'\neq n$, for different dataset placement configurations. To that end, we denote the set of variables included in the subfunction $f_n$ as
\begin{align}
\mathcal{I}_{f_n}\triangleq\{W_k\,|\,f_n
=\underset{k\in\mathcal{K}_{\mathcal{P}_n, d}}{\prod} W_k\},\quad\forall n\in [N]\ .
\end{align}

\begin{lem} {\bf (Increase in joint influence due to summation.)}
\label{inf_sum}
 Let $f=f_n\oplus f_{n'}$, 
 where $n\neq n'$. 
 Consider two different subsets of datasets denoted by  
 $\mathcal{S}_1=\{W_k|W_k\in \mathcal{I}_{f_n}\}$ and $\mathcal{S}'_1=\{W_k|W_k\in \mathcal{I}_{f_n} \cup \mathcal{I}_{f_{n'}}\}$. 
 We then have:
 \begin{align}
 \label{eq:increase}
     {\rm Inf}_{\mathcal{S}'_1}(f)\ge{\rm Inf}_{\mathcal{S}_1}(f)\ .
 \end{align}

 \begin{proof}
     See Appendix \ref{App_B}.
 \end{proof}
 
\end{lem}

From Lemma~\ref{inf_sum}, the subsets with the lowest joint influence include datasets from the same $f_n$. We next derive 
a lower bound on the average joint sensitivity for the proposed setting using Lemmas~\ref{lem_inf_prod}-\ref{inf_sum}, and present the communication-optimal placement configuration in Theorem~\ref{achiv}.

\begin{theo} {\bf (
A communication-optimal placement configuration.)}
\label{achiv}
Given a $(K, N, M, \mathcal{S}, f)$ distributed computing scheme, the average joint sensitivity and the total number of transmissions are lower bounded by
\begin{align}
\label{eq:as_bound}
    &{\rm as_{\mathcal{S}}}(f)\ge {\rm as_{\mathcal{S}^*}}(f)=\frac{N}{2^{M-1}}\ ,\nonumber\\
    &T^{(\mathcal{S})}\ge T^{(\mathcal{S}^*)}=N \ ,
\end{align}
respectively, corresponding to $\mathcal{S}^*=\{\mathcal{S}^*_n\,|\,n\in [N]\}$, 
where $\mathcal{S}^*_n=\{W_k\,|\,W_k\in \mathcal{I}_{f_n}\}$.


\begin{proof}
    See Appendix \ref{App_C}.
\end{proof}

\end{theo}

\begin{cor}
\label{cor_tr}
The number of transmissions by server $n$ is a monotonically increasing function in terms of the joint influence of datasets in each subset on Boolean function $f$, i.e., $|Z_n|\leq|Z_{n'}|$ if and only if ${\rm Inf}_{\mathcal{S}_n}(f)\leq{\rm Inf}_{\mathcal{S}_{n'}}(f),\,\forall n\neq n'$. 
\end{cor}

From Corollary \ref{cor_tr}, it is easy to observe that $T^{(\mathcal{S})}$ 
is a monotonically increasing function of the average joint sensitivity, i.e., $T^{(\mathcal{S})}\leq T^{(\mathcal{S}')}$ if and only if ${\rm as}_{\mathcal{S}}(f)\leq {\rm as}_{\mathcal{S}'}(f)$.

\section{Conclusions}
\label{sec:conclusion}
In this work, using the concept of sensitivity and influence, we introduced a novel approach for determining the interplay between communication cost and placement configuration for distributed computing of Boolean functions. In particular, we specified the optimal placement configuration from a communication cost perspective for a class of linearly-separable Boolean functions. Our approach is based on 
grouping the datasets to minimize the summation of their joint influences. As a future direction, we will extend our approach to distributed computing of nonlinear Boolean functions.



\appendix
\subsection{Proof of Lemma \ref{lem_inf_prod}}
\label{App_A}
    Using Definition \ref{joint_inf} and its probabilistic nature, we have:
     \begin{align}
        {\rm Inf}_{\mathcal{S}_n}(f)&= \mathbb{P}\Big[f(\mathbf{W}\oplus e_{\mathcal{S}_n})\neq f(\mathbf{W})\Big]\nonumber\\
        &\overset{(*)}{=} \frac{2}{2^{d}}=\frac{1}{2^{d-1}}\ 
 ,\nonumber
    \end{align}
    where $(*)$ holds since only two sequences $11\ldots 1$ and  $00\ldots 0$ out of $2^d$ possible sequences are feasible. Similarly, for the individual variable $W_k$, the influence is calculated as 
    \begin{align}
        {\rm Inf}_{k}(f)=\mathbb{P}\Big[\underset{i\in \mathcal{K}_{\mathcal{P}\backslash k, d-1}}{\prod} W_i\neq 0\Big]
        =\Big(\frac{1}{2}\Big)^{d-1}\ ,
    \end{align}
   where the last step follows when all the $d-1$ datasets $W_k\sim {\rm Bern}(\frac{1}{2})$, $k \in \mathcal{K}_{\mathcal{P}\backslash k, d-1}$ are equal to $1$. Therefore, (\ref{eq_inf_prod}) holds.

\subsection{Proof of Lemma \ref{inf_sum}}
\label{App_B}
According to Lemma \ref{lem_inf_prod}, ${\rm Inf}_{\mathcal{S}_1}(f)=\frac{1}{2^{d-1}}$.
To find the joint influence of subset $\mathcal{S}'_1$ on $f=f_n\oplus f_{n'}$, we first consider only one dataset difference between $\mathcal{S}_1$ and $\mathcal{S}'_1$, i.e., we assume $\mathcal{S}'_1=\{W_k\vert W_k\in \mathcal{I}_{f_n}\backslash W_n \cup W_{n'}\}$. We then decompose the respective product subfunctions as 
$$f_j=\underset{k\in\mathcal{K}_{\mathcal{P}_j \backslash j, d-1}}{\prod} W_k W_j \ ,\quad j\in \{n, \, n'\}\ .$$

Using Definition~\ref{joint_inf}, we can rewrite ${\rm Inf}_{\mathcal{S}'_1}(f)$ as
\begin{align}
\label{eq_inf_sum}
    {\rm Inf}_{\mathcal{S}'_1}(f)&=\mathbb{P}[\underset{k\in\mathcal{K}_{\mathcal{P}_n \backslash n, d-1}}{\prod} (W_k\oplus 1) W_n\nonumber\\
    &\oplus \underset{k\in\mathcal{K}_{\mathcal{P}_{n'} \backslash n', d-1}}{\prod} W_k (W_{n'}\oplus 1)\nonumber\\
    &\neq\underset{k\in\mathcal{K}_{\mathcal{P}_n \backslash n, d-1}}{\prod}
    W_k W_n\oplus \underset{k\in\mathcal{K}_{\mathcal{P}_{n'} \backslash n', d-1}}{\prod} W_k W_{n'}]\nonumber\\
    &=\mathbb{P}\Big[\bigoplus_{i=0}^{d-2}\underset{k\in\mathcal{K}_{\mathcal{P}_n \backslash n, i}}{\prod} W_k W_n \oplus \underset{k\in\mathcal{K}_{\mathcal{P}_{n'} \backslash n', d-1}}{\prod} W_k \neq 0\Big]\ .
\end{align}
Let $\Delta f\triangleq\bigoplus_{i=0}^{d-2}\underset{k\in\mathcal{K}_{\mathcal{P}_n \backslash n, i}}{\prod} W_k W_n \oplus \underset{k\in\mathcal{K}_{\mathcal{P}_{n'} \backslash n', d-1}}{\prod} W_k$. Considering the law of total probability and i.i.d. datasets:
\begin{align*}
\mathbb{P}[\Delta f=1]&=\frac{1}{2} \left(\mathbb{P}[\Delta f=1\,|\,W_n=1]\right.\nonumber\\
&\left.+\mathbb{P}[\Delta f=1\,|\,W_n=0]\right)\ ,
\end{align*}
where using Lemma \ref{lem_inf_prod}, we obtain:
\begin{align*}
\mathbb{P}[\Delta f=1\,|\,W_n=0]&=\mathbb{P}[\underset{k\in\mathcal{K}_{\mathcal{P}_{n'} \backslash n', d-1}}{\prod} W_k=1]\nonumber\\
&=\Big(\frac{1}{2}\Big)^{d-1} \ ,
\end{align*}
\begin{align}
\label{eq_inf_s1p}
&\mathbb{P}[\Delta f=1\,|\,W_n=1]\nonumber\\
&=\mathbb{P}\Big[\bigoplus_{i=0}^{d-2}\underset{k\in\mathcal{K}_{\mathcal{P}_n \backslash n, i}}{\prod} W_k \oplus \underset{k\in\mathcal{K}_{\mathcal{P}_{n'} \backslash n', d-1}}{\prod} W_k=1\Big]\nonumber\\
&=\mathbb{P}\Big[\bigoplus_{i=0}^{d-2}\underset{k\in\mathcal{K}_{\mathcal{P}_n \backslash n, i}}{\prod} W_k =0, \underset{k\in\mathcal{K}_{\mathcal{P}_{n'} \backslash n', d-1}}{\prod} W_k=1\Big]\nonumber\\
&+\mathbb{P}\Big[\bigoplus_{i=0}^{d-2}\underset{k\in\mathcal{K}_{\mathcal{P}_n \backslash n, i}}{\prod} W_k =1, \underset{k\in\mathcal{K}_{\mathcal{P}_{n'} \backslash n', d-1}}{\prod} W_k=0\Big]\nonumber\\
&=\Big[1-\Big(\frac{1}{2}\Big)^{d-2}\Big] \Big(\frac{1}{2}\Big)^{d-1} +\Big(\frac{1}{2}\Big)^{d-2} \Big[1-\Big(\frac{1}{2}\Big)^{d-1}\Big]\ .
\end{align}

After rearranging \eqref{eq_inf_s1p} and applying it to \eqref{eq_inf_sum}, we obtain:
\begin{align}
    {\rm Inf}_{\mathcal{S}'_1}(f)=2 \Big(\frac{1}{2}\Big)^{d-1} \Big[1-\Big(\frac{1}{2}\Big)^{d-1}\Big]\ .
\end{align}


It is obvious that ${\rm Inf}_{\mathcal{S}'_1}(f)\ge(\frac{1}{2})^{d-1}={\rm Inf}_{\mathcal{S}_1}(f)$. By induction, it follows that for any subset $\mathcal{S}'_1$ with more difference than one dataset compared to $\mathcal{S}_1$, \eqref{eq:increase} holds.

\subsection{Proof of Theorem \ref{achiv}}
\label{App_C}
We prove this theorem in two parts:
\paragraph{Achievability} For distributed computing of a linearly-separable Boolean function $f$ of degree $M$, there exists an achievable scheme with a placement configuration $\mathcal{S}^*$.

According to Lemma \ref{lem_inf_prod}, ${\rm Inf_{\mathcal{S}_n^*}}(f)=\frac{1}{2^{M-1}}, \forall n\in[N]$. 
The average joint sensitivity in this case is therefore:
\begin{align}
    {\rm as_{\mathcal{S}^*}}(f)&=\sum_{n=1}^N \frac{1}{2^{M-1}}\nonumber\\
    &=\frac{N}{2^{M-1}}\ .
\end{align}

According to Definition~\ref{ach}, the $(K, N, M, \mathcal{S}^*, f)$ distributed computing scheme is achievable since the user with cache configuration $\mathcal{S}^*$ would be able to recover the function in an error-free manner with only summing $Z^*_n$'s together, i.e.,
\begin{align}
    D(Z^*)&=\bigoplus_{n=1}^N Z^*_n\nonumber\\
    &= \bigoplus_{n=1}^N \underset{k\in \mathcal{K}_{\mathcal{P}^*_n , M}}{\prod} W_k\nonumber\\
    &=\bigoplus_{n=1}^N f_n=f\ .
\end{align}

\paragraph{Optimality (converse)}

Based on Lemma \ref{inf_sum}, the minimum joint influence happens when we group the datasets from the same product subfunction. We then use it to show the optimality of our proposed placement configuration.

We then examine a placement scheme $\mathcal{S}$, where we consider similar placement as $\mathcal{S}^*$ for $N-2$ servers.
For simplicity, we only swap two datasets between the two first servers. They will therefore contain subsets $\mathcal{S}_1=\{W_k\,|\,W_k\in \mathcal{I}_{f_n}\backslash W_n \cup W_{n'}\}$ and $\mathcal{S}_2=\{W_k\,|\,W_k\in \mathcal{I}_{f_{n'}}\backslash W_{n'} \cup W_n\}$, respectively.
In other words, we group $M-1$ datasets from $\mathcal{I}_{f_n}$ and one variable ($W_{n'}$) from $\mathcal{I}_{f_{n'}}$ in $\mathcal{S}_1$ and vice versa in $\mathcal{S}_2$.
According to Lemma \ref{lem_inf_prod} and Lemma \ref{inf_sum}, we have:
\begin{align}
    {\rm Inf}_{\mathcal{S}_n}(f)\ge{\rm Inf}_{\mathcal{S}^*_n}(f)=\frac{1}{2^{M-1}},\ \forall n=1,2\ .
\end{align}


For other $N-2$ subsets, we also have:
\begin{align}
    {\rm Inf}_{\mathcal{S}_n}(f)={\rm Inf}_{\mathcal{S}^*_n}(f)=\frac{1}{2^{M-1}}\ .
\end{align}

Using Definition~\ref{def:ajs} and summing the joint influences together for both $\mathcal{S}$ and $\mathcal{S}^*$ completes the proof. By induction, it follows that for any placement configuration $\mathcal{S}$ with more swapped datasets between subsets compared to $\mathcal{S}^*$, \eqref{eq:as_bound} holds. We also note that the minimum value of $T^{(\mathcal{S})}$ corresponds to $\mathcal{S}^*$ and equals to $N$, i.e., the transmission of $N$ units of data.

\Urlmuskip=0mu plus 1mu\relax
\bibliographystyle{IEEEtran}
\bibliography{Derya}

\end{document}